\documentclass[aps,pre,twocolumn,superscriptaddress,showpacs]{revtex4}

\usepackage{amsfonts,amssymb,amsmath,latexsym,epsfig,wasysym,bbold} 
\usepackage[sort&compress]{natbib}
\usepackage{float}
\usepackage{placeins}

\newcommand{\ket}[1]{\left| #1 \right\rangle}
\newcommand{\bra}[1]{\left\langle #1 \right|}

\usepackage{color,xcolor}
\usepackage{framed}
\colorlet{shadecolor}{green!15}

\begin{document}

\title{Quantum signatures of classical multifractal measures}
\author{Moritz Sch\"onwetter}
\author{Eduardo G. Altmann}
\affiliation{Max Planck Institute for the Physics of Complex Systems, 01187 Dresden, Germany}

\begin{abstract} 
A clear signature of classical chaoticity in the quantum regime is the fractal Weyl law, which connects the density of eigenstates to the dimension $D_0$ of the classical invariant set of open systems. 
Quantum systems of interest are often {\it partially} open (e.g., cavities in which trajectories are partially reflected/absorbed).
In the corresponding classical systems $D_0$ is trivial (equal to the phase-space dimension), and the fractality is manifested in the (multifractal) spectrum of R\'enyi dimensions $D_q$.
In this paper we investigate the effect of such multifractality on the Weyl law.
Our numerical simulations in area-preserving maps show for a wide range of configurations and system sizes $M$ that (i) the Weyl law is governed by a dimension different from $D_0=2$ and (ii) the observed dimension oscillates as a function of $M$ and other relevant parameters.
We propose a classical model which considers an undersampled measure of the chaotic invariant set, explains our two observations, and predicts that the Weyl law is governed by a non-trivial dimension $D_\mathrm{asymptotic} < D_0$ in the semi-classical limit $M\rightarrow\infty$.
\end{abstract} 


\maketitle

\section{Introduction}

Fractality is a well-known signature of chaotic dynamics~\cite{Ott:Book2002,Lai:Book2011,Aguirre:RMP2009}. 
In open quantum systems the fractality of the classical dynamics appears in the distribution of eigenstates~\cite{Casati:PD1999} and in the Weyl law~\cite{Lu:PRL2003}.
Recent studies considered the effect of periodic orbits~\cite{Pedrosa:PRE2012} and weak chaos~\cite{Ramilowski:PRE2009,Körber:PRL2013,Schomerus:PRL2004} and were conducted on $4$-dimensional Hamiltonian systems~\cite{Ramilowski:PRE2009} and different area-preserving maps~\cite{Pedrosa:PRE2012,Novaes:JPA2013,Körber:PRL2013,Schomerus:PRL2004}.
In these systems the fractal Weyl law can be written as
\begin{equation}
   \label{eq:def:fwl}
   N(|\nu_i|>|\nu|_\mathrm{cutoff})\sim M^{D_0^\mathrm{(S)}/2},
\end{equation}
where $M$ is the dimension of the Hilbert space, \(D_0^\mathrm{(S)}\) is the fractal dimension of the chaotic saddle, and \(|\nu|_\mathrm{cutoff}\) is a cutoff separating long-lived from short-lived states with eigenvalues $\nu_i$.

Optical cavities~\cite{Wiersig:PRE2008} and other relevant physical systems~\cite{Altmann:RMP2013} are not completely open but instead show partial absorption.  
A simple physical picture is a billiard in which trajectories hitting the boundary are partially reflected and partially absorbed.
In contrast to open systems (full absorption),  in {\it partially} absorbing systems, the box-counting dimension $D_0$ is an integer number equal to the phase-space dimension~\cite{Altmann:PRL2013}. 
Fractality in the classical dynamics of such systems is still present~\cite{Wiersig:PRE2008} and can be quantified using the spectrum of R\'enyi dimensions~\cite{Ott:Book2002}
\begin{equation}\label{eq:def:renyi_dimensions}
   D_q=\frac{1}{1-q}\lim_{\varepsilon\rightarrow 0}\frac{\ln\sum_{i=1}^{N{(\varepsilon)}}\mu_i^q}{\ln 1/\varepsilon},
\end{equation}
where ${\mu\left(\vec{x}\right)}$ is the relevant probability measure (${\int_{\Omega}\mathrm{d}\mu\left(\vec{x}\right)\equiv1}$) in the phase-space $\Omega$ and the sum is over $\varepsilon$-sized boxes \(B_i\) (with $0<\mu_i:=\int_{B_i}\mu\left(\vec{x}\right)\mathrm{d}\vec{x}$) used to partition the phase-space.
For $q=0$,  \(D_0\) is recovered.
A distribution ${\mu\left(\vec{x}\right)}$ with a non-trivial dependence of $D_q$ on $q$ is called \emph{multifractal}.  

In this paper we investigate the quantum mechanical signatures of multifractality in the classical phase-space. 
While a straightforward application of the fractal Weyl law to partially open systems predicts a trivial scaling ($N \sim M$), as also argued in~\cite{Nonnenmacher:PRE2008}, non-trivial scalings have been reported in numerical investigations~\cite{Wiersig:PRE2008}. 
In order to investigate the robustness of fractality in the Weyl law we study it at the transition between monofractal and multifractal.
This is done by varying the reflectivity $R\ge0$ of an absorbing region localized in the phase-space. 
For all nonzero reflectivities we obtain non-trivial Weyl laws with effective dimensions that show strong oscillations as a function of system size and other parameters.

We argue that these observations are due to an effective undersampling of the classical measure in the quantum regime and that the non-trivial scaling remains valid in the semiclassical limit $M\rightarrow\infty$.

\begin{figure*}[]
   \centering
   \includegraphics[width=\textwidth]{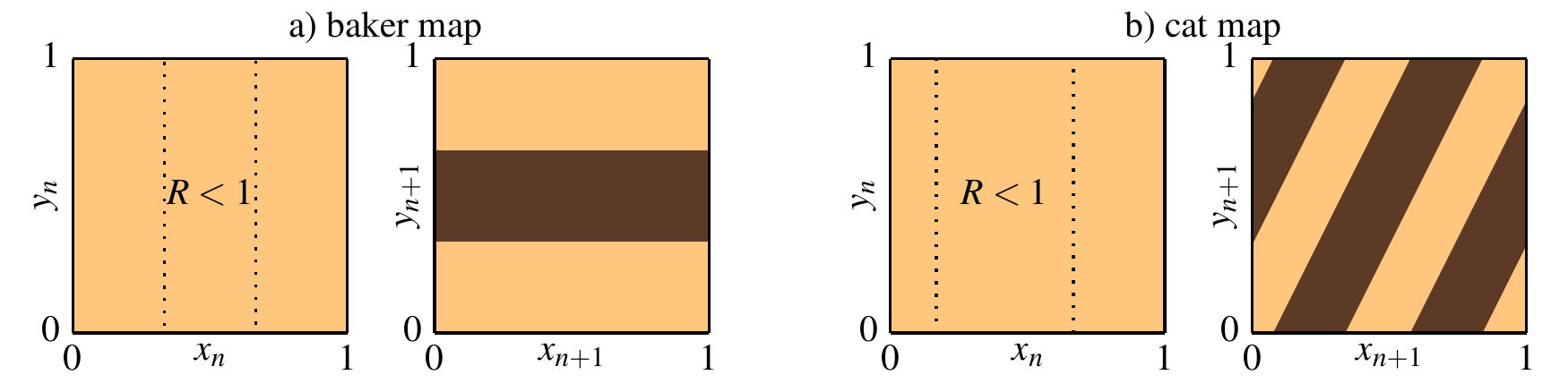}
   \caption
      {
         (Color online) 
         Classical dynamics of area-preserving maps $(x_n,y_n)\mapsto(x_{n+1},y_{n+1})$ with absorbing regions. 
         The pairs of plots a) and b) show the action of one application of the baker's map ${(x_{n+1},y_{n+1})=(3x-\lfloor 3x_n \rfloor, y_n/3 + \lfloor 3y_n \rfloor/3)}$ and the cat map ${(x_{n+1},y_{n+1})=(2x_n+y_n\mod1,3x_n+2y_n\mod1)}$, respectively.
         The areas surrounded by dotted lines in the left plots indicate the absorbing regions ($R<1$) of each map.
         The dark regions in the right plots show the first images of the absorbing region.
      }
   \label{fig:sketches}
\end{figure*}

\section{Fractals in Classical Dynamics}\label{sec:classical_dynamics}

As numerically convenient examples of Hamiltonian chaotic systems we consider here fully chaotic symplectic maps of the form ${f:\vec{x}_n\mapsto\vec{x}_{n+1}}$, defined in a bounded phase-space $\vec{x}\in\Omega$.
To incorporate the effect of partial absorption we associate to each trajectory an intensity $I_{n}$, with $I_0=1$, $I_{n+1}=R(\vec{x}_n) I_n$, and $R(\vec{x})\in\left[0,1\right]$ is a real valued function describing the reflectivity of different regions in the phase-space. 

In order to investigate the differences between partially and fully open systems it is
instructive to consider the special case of ${R(\vec{x})=R<1}$ in a small absorbing region of the phase-space (i.e. ${R(\vec{x})=1}$ everywhere else).
If $R=0$ in the absorbing region the system corresponds to usual open maps and the set of initial conditions $\vec{x}_0$ with $\lim_{n\rightarrow\infty}I_{n}=1$ form the forward trapped set  of the map (the stable manifold of the chaotic saddle~\cite{Lau:PRL1991,Lai:Book2011,Jung:JPA1991}).
In this case the measure is uniformly distributed on a fractal support and the R\'enyi spectrum is flat ($D_q=D_0$).
In analogy, for a partially absorbing system $0<R<1$ the measure in a phase-space region $B_i$ (e.g., the \(i\)-th box of a partition by a regular grid) should be proportional to the average intensity $I_{n}$ for $n\rightarrow \infty$ of randomly drawn $\vec{x}_{0}\in B_i$.
The finite time  normalised measure of $B_i$ after $n$ iterations can thus be written as
\begin{equation}\label{eq:mu}
   \mu_{n,i}=\frac{\langle I\rangle _{n,i}}{\sum_{i}\langle I \rangle_{n,i}},
\end{equation}
where the average intensity $\langle ... \rangle_{n,i}$ is computed over initial conditions $\vec{x}_0\in B_i$.
This definition is consistent with the usual definition of the stable manifold measure~\cite{Ott:Book2002} considering the intensities $I\in\left[0,1\right]$ of trajectories as weights (see also Ref.~\cite{Altmann:RMP2013}).

\begin{figure}[]
\includegraphics[width=0.5\textwidth]{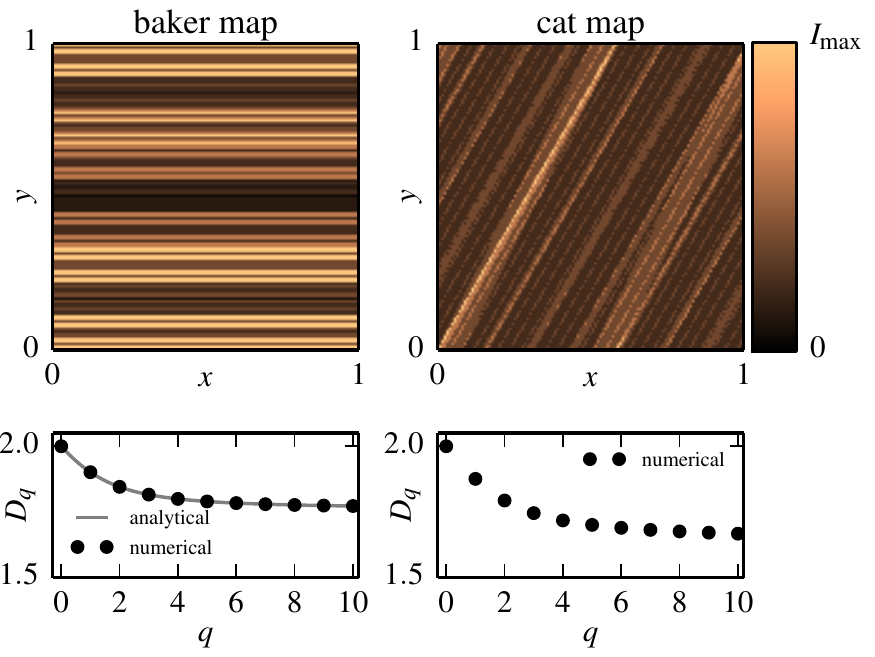}
\caption{
   (Color online) 
   Multifractal properties of the maps defined in Fig.~\ref{fig:sketches}.
   {\bf Top row}: finite time approximation of a typical phase-space distribution of the measure $\mu_i$ defined in Eq.~\eqref{eq:mu}.   
   {\bf Bottom row}: R\'enyi spectrum $D_q$ for  $R=0.3$ (see Appendix~\ref{ap:baker} for the analytical calculations on the baker map).
}
\label{fig:dq_and_measures}
\end{figure}

As representative cases of fully chaotic area-preserving maps we investigate in detail the baker~\cite{Ott:Book2002} and the cat map~\cite{Dematos:AnnPhys1995} as defined in Fig.~\ref{fig:sketches}.
In the case of the baker map the absorbing region was chosen in order to allow for comparison with analytical calculations.
In the case of the cat map we chose an arbitrary asymmetric region.
The essential point for our investigation of the role of partial opening is that both choices lead to a non-trivial $D_0$ for $R=0$.
In Fig.~\ref{fig:dq_and_measures} we report the distribution of the measure $\mu$ in the phase-space for a particular reflectivity and the spectrum $D_q$.
For the baker map the calculations can be done analytically (see Appendix~\ref{ap:baker}) and confirm the validity of our numerical methods of estimating $\mu_i$ and $D_q$.
In both maps $D_0=2$ and $D_q$ is non-trivial, confirming the result of Ref.~\cite{Altmann:PRL2013} that the apparent self-similar distribution of $\mu$ observed in the phase-space is properly quantified only through the full multifractal spectrum $D_q$. 

\section{Fractals in Quantum Systems}\label{sec:fractal_in_qd}

In this section we introduce the quantised version of the maps of Sec.~\ref{sec:classical_dynamics} and compute Weyl's law~(\ref{eq:def:fwl}).
By \emph{open quantum map} we mean a propagator \cite{Wisniacki:PRE2008}
\begin{equation}
   \hat{\mathcal{M}}=\hat{\mathcal{U}}\hat{\mathcal{P}},
\end{equation}
where \(\hat{\mathcal{U}}\) is the quantised closed system and \(\hat{\mathcal{P}}\) is a projector that contains all the information about absorption.
In the case of full absorption, \(\hat{\mathcal{P}}\) removes the part of the quantum wave-function \(\psi(x)\) that overlaps with the absorbing region, but does not alter the remaining part.
In the case of partial absorption, \(\hat{\mathcal{P}}\) reduces the intensity by a factor $R(x)$ as \({\left|\bra{x}\hat{\mathcal{P}}\ket{\psi}\right|^2=R(x)\left|\psi(x)\right|^2}\).
See App.~\ref{sec:app:quantum_maps} for the matrix representations of the maps used in this paper. 

In an \(M\)-dimensional Hilbert space (effective \(\hbar=\frac{1}{2\pi M}\))  maps $\hat{\mathcal{M}}$ have have \(M\) left (and right) eigenstates with complex energies \(\epsilon_i=E_i+\mathrm{i}\Gamma_i/2\).
$\hat{\mathcal{M}}$ acts like a time evolution operator so its eigenvalues \(\nu_i\) are related to  \(\epsilon_i\) by
\begin{equation}\label{eq:def:nu}
   \nu_i=\exp\left[-\frac{\mathrm{i}}{\hbar}\epsilon_i\right]=\exp\left[-\frac{\mathrm{i}}{\hbar} E_i\right]\exp\left[-\frac{1}{2\hbar}\Gamma_i\right].
\end{equation}
Since \(\Gamma_i\in\mathbb{R}^+_0\) and \(E_i\in\mathbb{R}\) it follows that \(\left|\nu_i\right|\in\left[0,1\right]\). Hereafter we use $\left|\nu_i\right|$ to quantify the lifetime of the state, e.g. long-lived states have $\left|\nu_i\right| \lessapprox 1$ (\(\Gamma_i\ll1\)).  

\begin{figure}[]
\includegraphics[width=1\columnwidth]{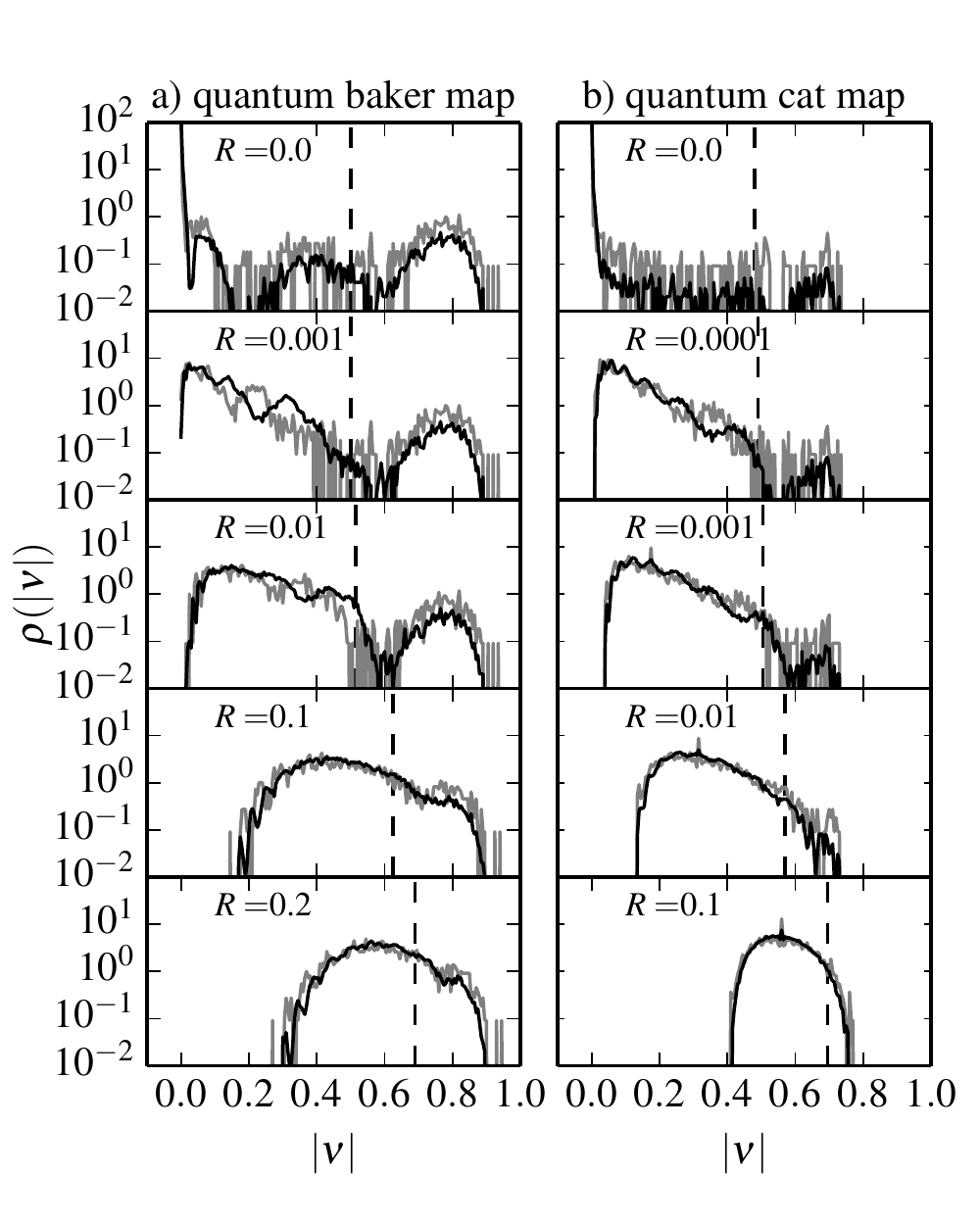}
\caption
   {
   Distribution of eigenvalues $|\nu|$, see Eq.~\eqref{eq:def:nu})in the quantised baker (a) and cat (b) maps.
   Two matrix sizes were used:  $M=3^{7}$ (gray line)  and $M=3^9$ (black line)~\cite{foot1}.
   The vertical dashed line marks the cutoff value used for the fractal Weyl law calculations in Fig.~\ref{fig:weyl}ad.
   }
\label{fig:eigenvalues}
\end{figure}

We calculated the eigenvalues of \(\hat{\mathcal{M}}\)  for both the baker and cat map for different reflectivities \(R\) and matrix sizes \(M=3^n\) with $n\in\mathbb{N}$~\cite{foot1}.
The normalised distributions of \(\left|\nu_i\right|\) are shown in Fig.~\ref{fig:eigenvalues}.
The states concentrate around a typical value $\left|\nu\right|_{\mathrm{typ}}$, which is $\left|\nu\right|_{\mathrm{typ}}\approx0$ for $R=0$, increases with $R$, and can be approximated as the median of $\rho(\left|\nu\right|)$.
For increasing matrix-sizes $M$ the distribution becomes more concentrated around~$\left|\nu\right|_{\mathrm{typ}}$.
To investigate this tendency in further detail, and in line with the fractal Weyl law, we concentrate on the number of eigenstates with $\left|\nu_i\right|\geq\left|\nu\right|_\mathrm{cutoff}$.

\begin{figure*}[]
\includegraphics[width=1\textwidth]{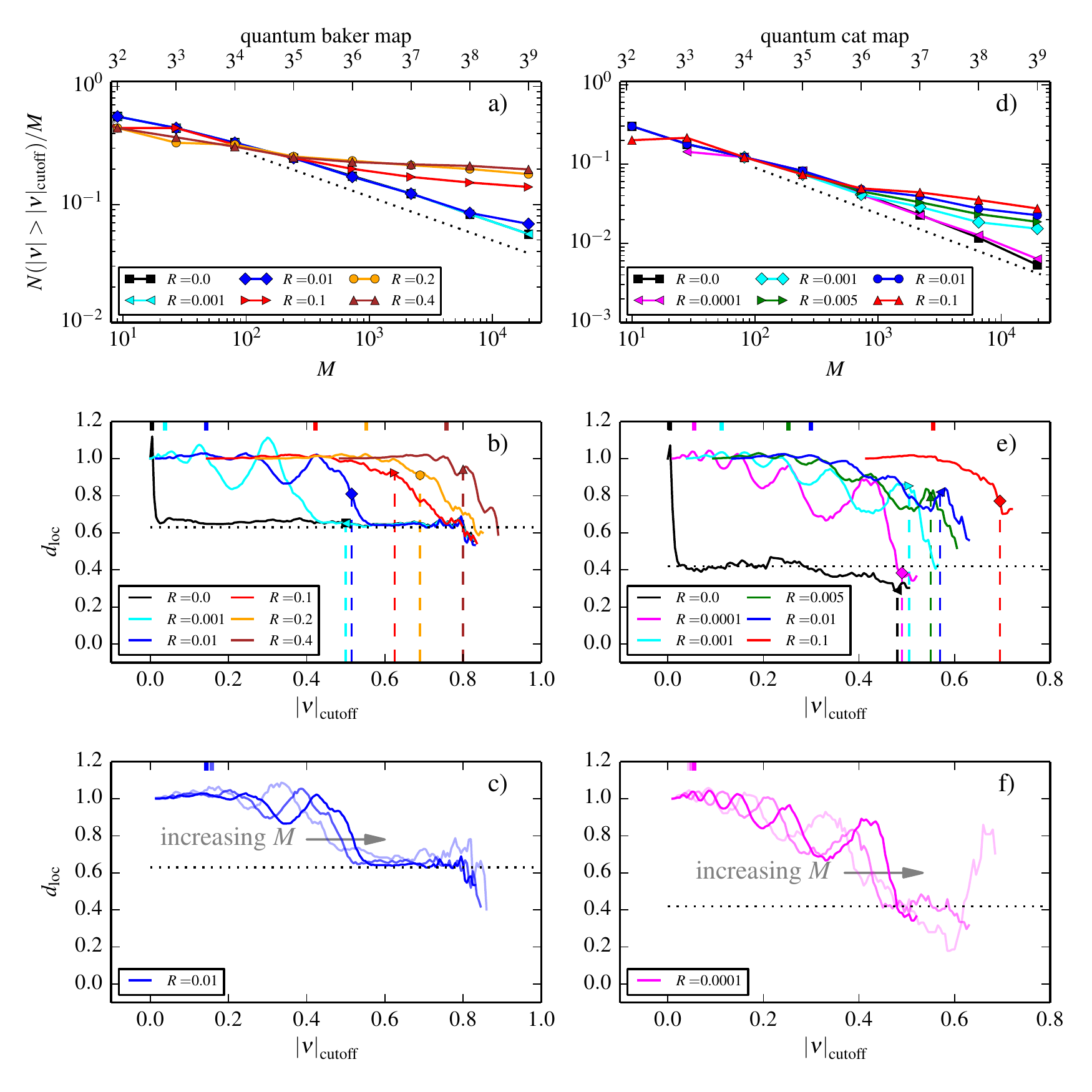}
\caption
   {
      (Color online)
      Scaling behaviour of the quantum states for the baker (a-c) and cat (d-f) maps.
      \textbf{Top row}: Fraction of eigenvalues above the cutoff value $\left|\nu\right|_\mathrm{cutoff}$ as a function of the system size \(M\) at different reflectivities.
      The dotted lines mark the expected scaling for \(R=0\) (\(d_0^{(R=0)}=\ln(2)/\ln(3)\) for the baker map and $d_0^{(R=0)}\approx0.42$ for the cat map).
      \textbf{Middle row}: Local scaling exponent \(d_\mathrm{loc}(M):=\left(\ln N(M)-\ln N(M/3)\right)/\ln3\), where \(N(M)\) denotes \({N(\nu_i:\left|\nu_i\right|\geq\left|\nu\right|_\mathrm{cutoff})}\), at \(M=3^9\) for different values of $R$ as a function of $\left|\nu\right|_\mathrm{cutoff}$ (colors and symbols as in top row).
      We show results up to a cutoff where \(N(M)<0.005M\) after which point the lack of resonances leads to strong oscillations of \(d_\mathrm{loc}\).
      The vertical dashed lines indicate the cutoff values used in the top row.
      The small ticks on the upper axis mark the positions of the medians of the eigenvalue distributions ($\approx \left|\nu\right|_{\mathrm{typ}}$).
      \textbf{Bottom row}: \(d_\mathrm{loc}(M)\) as a function of the cutoff for $M=\{3^7,3^8,3^9\}$ (lighter shades correspond to smaller $M$).
   }
\label{fig:weyl}
\end{figure*}

Figures~\ref{fig:weyl}ad show the \emph{relative} number of eigenstates \({N(\nu_i:\left|\nu_i\right|\geq\left|\nu\right|_\mathrm{cutoff})/M}\) as a function of the matrix-size~$M$. 
For this plot, $\left|\nu\right|_\mathrm{cutoff}$ was  chosen in such a way that $\left|\nu\right|_\mathrm{cutoff}>\left|\nu\right|_{\mathrm{typ}}$ but still small enough to allow the calculation of the scaling for a wide range of $M$, see vertical dashed lines in Fig.~\ref{fig:eigenvalues}~\footnote{In practice, for each of the maps the number of states with ${\left|\nu_i\right|\geq\left|\nu\right|_\mathrm{cutoff}}$ was chose to be the same for all reflectivities at $M=243$ (82 for cat-map): ${\left|\nu\right|_\mathrm{cutoff}=\min\left\{\left|\nu\right|_c:N(\nu_i:\left|\nu_i\right|\geq\left|\nu\right|_c)>N_c \right\}}$ where \({N_c=60}\) for the baker map and  \({N_c=10}\) for the cat map.}.
The division by $M$ allows for an improved analysis of the scaling, which according to the Weyl law prediction~\eqref{eq:def:fwl} should be $d_0-1$, where \(d_0=D_0-1\) is the partial fractal dimension of the forward trapped set along the unstable direction.
\(D_0\) denotes the full fractal dimension of this set.
It is connected to the dimension of the chaotic saddle \(D_0^\mathrm{(S)}\) by~\cite{Lai:Book2011} 
\begin{equation}
   D_0^\mathrm{(S)}=2D_0-2\quad\Leftrightarrow\quad d_0=\frac{D_0^\mathrm{(S)}}{2}.
   \label{eq:partial_dimensions}
\end{equation}
The usual fractal Weyl law is confirmed in Fig.~\ref{fig:weyl}ad for the case \(R=0\).
For small $R$, the curves follow the $R=0$ case for small $M$ before deviating from it. 
With increasing $R$,  the deviation is observed already for smaller $M$.
Interestingly, even for large $R$ and $M$ the numerical results do not approach a constant value as predicted by the classical dimension $d_0=1$. 
Instead, a non-trivial scaling with an {\em effective} dimension between 0 and 1 is observed.
Even  if the numerical results do not allow for a conclusion regarding the scaling in the limit $M\rightarrow\infty$, the roughly constant scaling for a broad range of values in $M$ shows the importance of the effective dimension (similar effective dimensions have been employed to investigate classical~\cite{Motter:PRE2005} and quantum~\cite{Körber:PRL2013} Hamiltonian systems with mixed phase-space). 

In order to more carefully analyse the scaling we define the \emph{local} dimension $d_{\mathrm{loc}}$ as the slope of the \(N/M\) vs. \(M\) curves between two nearby matrix sizes \(M\).
Figures~\ref{fig:weyl}be show the dependency of $d_{\mathrm{loc}}$ on the choice of the cutoff $\left|\nu\right|_\mathrm{cutoff}$.
For small cutoffs \(\left|\nu\right|_\mathrm{cutoff}< \left|\nu\right|_{\mathrm{typ}}\) we observe $d_{\mathrm{loc}}=1$, in agreement with the observation in Fig.~\ref{fig:eigenvalues} that the distribution of eigenvalues becomes concentrated around $\left|\nu\right|_{\mathrm{typ}}$ (almost all states count as long-lived).
A more interesting behavior appears for  \(\left|\nu\right|_\mathrm{cutoff} > \left|\nu\right|_{\mathrm{typ}}\), where at least two regimes governed by non-trivial $d_{\mathrm{loc}}$ can be identified before $d_{\mathrm{loc}}$ starts to fluctuate for $\left|\nu\right|_{\mathrm{cutoff}}\rightarrow1$ due to a lack of statistics:
(I1) an regime of intermediate $\left|\nu\right|_{\mathrm{cutoff}}$ with strong (almost periodic) oscillations of $d_{\mathrm{loc}}$;
and (I2) a regime in which \(d_{\mathrm{loc}} \approx d_0^{(R=0)}\).
These regimes are particularly visible for small $R$ (because $\left|\nu\right|_{\mathrm{typ}} \rightarrow 0$), but can be identified also for larger $R$.
Surprisingly, Figs.~\ref{fig:weyl}cf show that regime I1 increases for increasing $M$ (strong oscillations cover a wider range of cutoffs).
This suggests that the oscillating (local) scaling  is not an artifact of small $M$ but instead that it will prevail in the large \(M\) limit.

In summary, our numerical observations of the quantum maps show surprising features.
In particular: (i) the classical prediction $d_0=1$ was not observed; (ii) the effective scaling varies smoothly with $R$; and (iii) strong oscillations we observed in $d_{\mathrm{loc}}$ vs. $\left|\nu\right|_{\mathrm{cutoff}}$.  

\section{Sampling Multi-Fractal Measures}

In this section we propose an explanation for the scaling of the number of long-lived
states with $M$ observed in the previous section based on the classical dynamics of the system. 

\subsection{General Argument}\label{sec:general_argument}

We first consider how the usual fractal Weyl law is obtained from the classical phase-space of fully absorbing maps~\cite{Lu:PRL2003,Schomerus:PRL2004}.
Lu {\it et  al.} \cite{Lu:PRL2003} argue that the states can be considered as non-overlapping and that each of them covers an area of \(h= 1/M\).
This localisation happens in the forward trapped set (for the left eigenstates \footnote{We focus on left eigenstates but the same ideas apply to the right ones, which localise in the backward trapped set}) \cite{Casati:PD1999,Keating:PRL2006}, which is smooth.
Therefore, the number of {\it different} long-lived states depends on the distribution of the forward trapped set in the perpendicular direction (along the backward trapped set or, equivalently, along the unstable manifold of the chaotic saddle).
The reasoning above leads to the following constructive classical procedure: 
\begin{itemize}
\item[1.] Cover an intersection of  the forward trapped set with segments of size $h =
  \varepsilon = 1/M$.
\item[2.] Count the number $N_b$ of boxes (segments) needed to cover the forward trapped set, which estimates the phase-space volume available to the long-lived eigenstates. 
\item[3.] The number of long-lived states $N(M)$ is proportional to $N_b$.
\end{itemize}
Since $N_b\sim M^{d_0}$ we obtain $N(M) \sim M^{d_0}$, the fractal Weyl's law~(\ref{eq:def:fwl}).
The scaling exponent \(d_0\) is the partial fractal dimension of the forward trapped set along the unstable direction~\eqref{eq:partial_dimensions}.
Schomerus and Tworzydlo~\cite{Schomerus:PRL2004} provide an alternative explanation based on the phase-space areas that do not escape up to a given time (which grows with $M$).
Both explanations are equivalent because asymptotically the $h$-sized boxes intersecting the invariant set correspond to the non-escaping phase-space areas.

We now generalize the ideas above to the case of partial absorption.
In this case we expect for large $M$ the long-lived states to be distributed according to the classical measure~$\mu$ of the forward trapped set.
Therefore, it is essential to consider a procedure that accounts for the non-uniformity of $\mu$.
In line with the usual interpretation of $\mu$ as a probability measure, we modify point 2. above and attribute to each segment $i$ a probability $p_i$ of being counted as long-lived (i.e., a probability that a long-lived state is localized in it).
This probability has to be large (small) where the measure \(\mu_i\) of the segment $i$ is large (small) because states localise on high measure areas of the phase-space.
At the same time, for large $M$ the number of possibly long-lived states grows and $p_i$ should grow because of the multiple chances of {\it one} state being localized in segment $i$.
The simplest stochastic process agreeing with these two intuitions corresponds to performing $S$ independent trials  with a success-probability \(\mu_i\). 
This can be incorporated in the procedure described above by replacing point 2. by 
\begin{itemize}
   \item [2'.] Count each segment $i$ as part of the accessible phase-space volume with a probability \({p_i=1-(1-\mu_i)^S}\), where \((1-\mu_i)^S\) is the probability that the box is rejected \(S\) times.
\end{itemize}
For one realization of this process the available volume $V(M)$ is proportional to the number $N_b$ of counted segments.
Therefore, point 3. is generalised to 3'. as 
\begin{equation}\label{eq:expected_value_general}
   N(M) \propto \mathrm{E}[N_b]=\sum_{i=1}^Mp_i=\sum_{i=1}^M1-(1-\mu_i)^S,
\end{equation}
where \(\mathrm{E}[...]\) is the expected value computed over multiple sampling realisations and the number of trials $S$ is associated to the total number of states~$S=S_0M$.
In the case of a fully absorbing region, \(\mu_i=1/\mathcal{N}\) for all \(\mathcal{N}\) boxes intersecting the forward trapped set and \(\mu_i=0\) everywhere else.
Eq.~\eqref{eq:expected_value_general} simplifies to \({\mathrm{E}[N_b]=\mathcal{N}\left(1-(1-1/\mathcal{N})^S\right)} \xrightarrow{S\rightarrow\infty} \mathcal{N}\) and we recover the fractal Weyl law.
Analogously, in the fully reflecting case $\mu_i=1/M$ and \({\mathrm{E}[N_b]\xrightarrow{S\rightarrow\infty}M}\) and we recover the traditional Weyl law. 

At the heart of the reasoning above is a crucial difference between the computation of $d_0$ and of the Weyl scaling. 
$d_0$ is obtained by first taking the limit of sampled trajectories $S$ to infinity $S\rightarrow\infty$ (construction of the measure) and \emph{afterwards} computing the scaling of filled boxes in the limit of small box sizes $\varepsilon\rightarrow0$.
In the Weyl law the analogues of these two limits ($S\rightarrow\infty$ and $\varepsilon\rightarrow0$) happen simultaneously when $M\rightarrow\infty$ because the maximum number of eigenstates grows with $M$ ($S \propto M$)  and the phase-space resolution increases with $M$ ($h = \varepsilon = 1/M$).
Therefore, the scaling in the Weyl law corresponds to the estimation of $d_0$ when sampling the measure with a sample size
\begin{equation}\label{eq:S:conjectured}
   S= S_0M = S_0 \varepsilon^{-1}
\end{equation}
in \eqref{eq:expected_value_general}.
The proportionality constant $S_0$ effectively controls how many segments (long-lived states) will be found for the different $M$ values and therefore plays a similar role as the cutoff value $\left|\nu\right|_{\mathrm{cutoff}}$ in the quantum case.

\subsection{Application to Baker Map}

We now apply the general sampling method proposed in the previous section to the baker map (defined in Fig.~\ref{fig:sketches}). 
The expected number of long-lived boxes $\mathrm{E}\left[N_b\right]$ for $\varepsilon=3^{-n}$ can be computed analytically (see Appendix \ref{sec:app:undersampling}) 
\begin{equation}\label{eq:expectation}
   \mathrm{E}\left[N_b\right]\\=3^n-\sum_{l=0}^n2^{n-l}\frac{n!}{l!(n-l)!}\left(1-\frac{R^l}{\left(2+R\right)^{n}}\right)^S.
\end{equation}

\begin{figure}[]
\includegraphics[width=.5\textwidth]{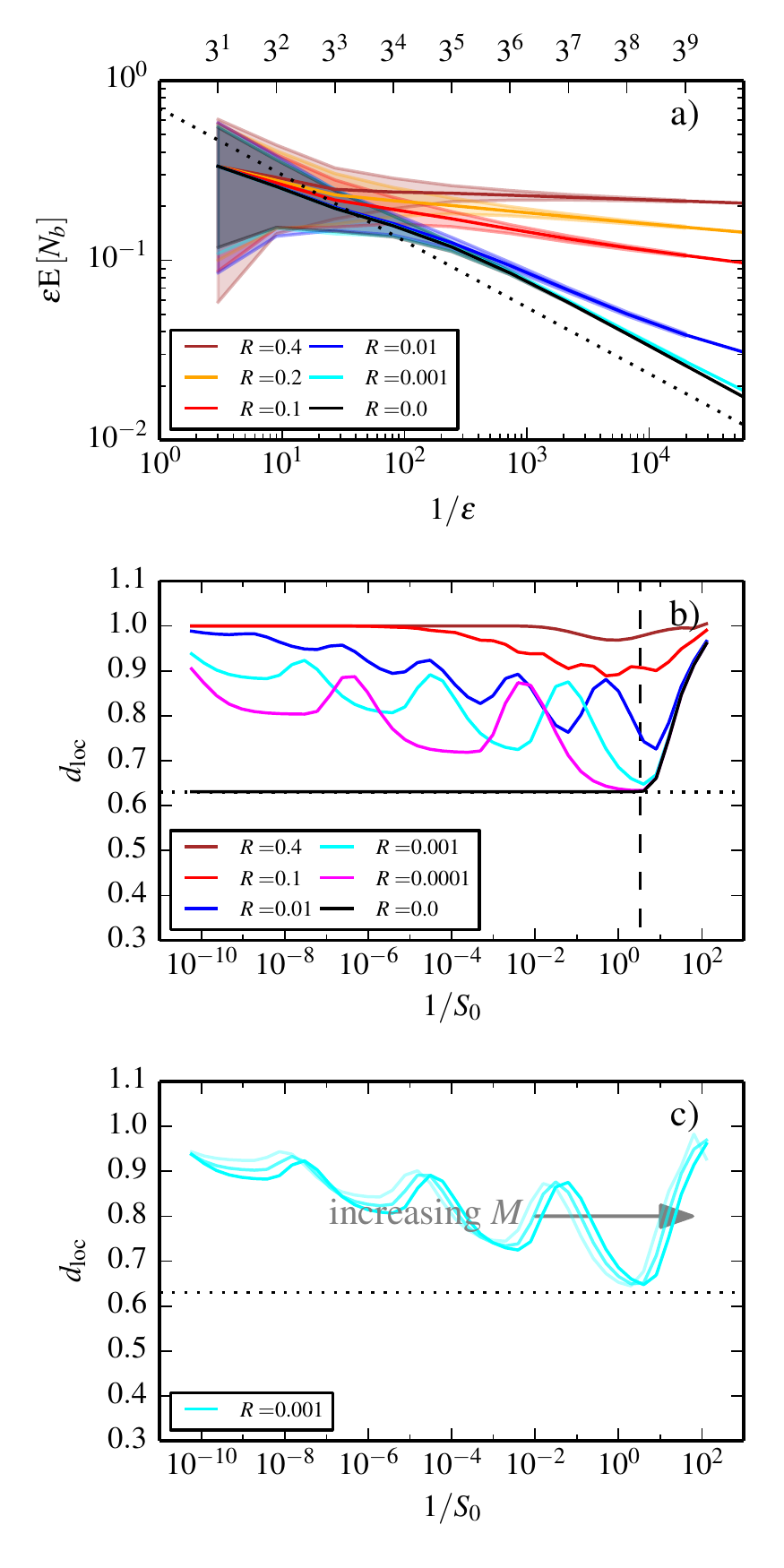}
\caption{
   (Color online)
   Scaling behaviour of the classical expected values \eqref{eq:expectation} for the baker map.
   The black dotted line denotes $d_0^{(R=0)}$.
   a) Rescaled expected value of accepted boxes at an arbitrary sample size of $S(M)=S_0M$ with $S_0=0.3$.
   \(\mathrm{E}[N_b]\) grows at fixed \(\varepsilon\) monotonically with \(R\).
   The shaded areas cover the range between \(\mathrm{E}[N_b]\pm\sigma[N_b]\), where \(\sigma\) is the standard deviation.
   b) Scaling exponent $d_\mathrm{loc}(M=3^9)$ as a function of $1/S_0$.
   The lower the value of $R$, the lower the curve intersects the dashed vertical line (the \(S_0\) choice used in a).
   c) $d_\mathrm{loc}(M=3^n)$ with \(n\in\{7,8,9\}\). Lighter shades correspond to smaller \(M\).
}
\label{fig:oscillations}
\end{figure}

In Fig.~\ref{fig:oscillations} we plot Eq.~(\ref{eq:expectation}) and its local derivatives as a function of its three parameters \(\varepsilon\), \(S_0\), and $R$.
The similarities to the quantum observations shown in Fig.~\ref{fig:weyl} are remarkable (the comparison is based on the identifications $\varepsilon = h =1/M$, ${N(\nu_i:\left|\nu_i\right|\geq\left|\nu\right|_\mathrm{cutoff})} \propto \mathrm{E}\left[N_b\right]$, and the similar role played by $S_0$ and $\left|\nu\right|_{\mathrm{cutoff}}$.
In particular, we reproduce the three main observations reported in Sec.~\ref{sec:fractal_in_qd}: 
   (i) the scaling of the curves is different from $d_0=1$;
   (ii) for increasing \(R\), the effective scaling deviates more and more from \(d_0^{(R=0)}\) and approaches a non-trivial dimension (Fig.~\ref{fig:oscillations}a);
   (iii) the local slope $d_{\mathrm{loc}}$ show strong oscillations with $S_0$  (Fig.~\ref{fig:oscillations}b) in a range of $S_0$ values which increases with $M$ (Fig.~\ref{fig:oscillations}c). 
One quantum feature which is not immediately apparent in Fig.~\ref{fig:oscillations}b is the plateau of \(d_\mathrm{loc}\approx d_0^{(R=0)}\), the regime I2 mentioned at the end of Sec.~\ref{sec:fractal_in_qd}.
To understand this we need to take a closer look at the mapping from \(S_0\) to the cutoff $\left|\nu\right|_{\mathrm{cutoff}}$.
The limiting case $S_0 \rightarrow \infty$ can be safely identified with $\left|\nu\right|_{\mathrm{cutoff}}\rightarrow 0$.
In this limit our model correctly reproduces the quantum observations ($d_{\mathrm{loc}}\rightarrow 1$).
The identification of the opposing limit $S_0 \rightarrow 0$ with $\left|\nu\right|_{\mathrm{cutoff}}\rightarrow 1$ is also possible but less meaningful, since in this limit both the number of resonances, and $N_b$ become very small, and as a result the determination of $d_\mathrm{loc}$ becomes prone to fluctuations. 
Between these limits, we can assume that $1/S_0$ grows monotonically with $\left|\nu\right|_{\mathrm{cutoff}}$.
However, the  functional form relating $1/S_0$ and $\left|\nu\right|_{\mathrm{cutoff}}$  is unknown (e.g., it depends on the distribution of eigenstates) which does not allow for a careful quantitative comparison of the two plots (e.g., a non-linear transformation of the x-axis in Fig.~\ref{fig:oscillations}b could generate the plateaus around $d_0^{(R=0)}$ seen in  Figs.~\ref{fig:weyl}bd.

\begin{figure}[]
\includegraphics[width=1\columnwidth]{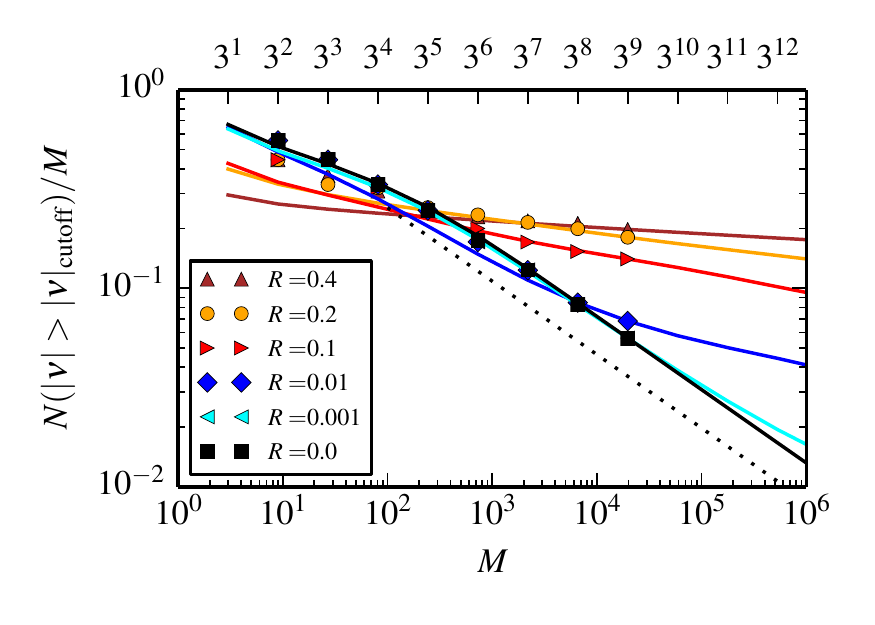}
\caption
   {
      (Color online)
      Comparison of the quantum data to the classical model.
      The symbols show the number of long-lived states.
      The solid lines are the classical values of \(a\mathrm{E}[N_b]\) obtained by choosing \(S_0\) and \(a\) for each \(R\) so that the rightmost two points of the quantum data are matched as close as possible.
  }
\label{fig:extrapolation}
\end{figure}

We now search for a quantitative comparison between our classical sampling model and the quantum observations.
The only parameter we have to fit is \(S_0\) in the sampling model, which according to the arguments of Sec .~\ref{sec:general_argument} should reproduce the {\it scaling} of the long-lived states with {\it large} \(M\).
Accordingly, for each value of \(R\) we choose the value of \(S_0\) which reproduces the local slope between the last two \(N/M\) points in the quantum data (Figs.~\ref{fig:weyl}ad) and we shift the \(\mathrm{E}\left[N_b\right]\) curve obtained to match the rightmost point of each quantum data-set (this last shift fixes the proportionality factor $a$ in $N(M)=a \mathrm{E}[N_b]$).
We restrict \(S_0\) to the range \(\left[3^{-1},3\right]\), so that \(S\) is of the order of \(M\) since these values give the best agreement with the whole range of quantum data.
Because of the oscillations in \(d_\mathrm{loc}\) there may be more than one choice of \(S_0\)~\footnote{The asymptotic behaviour is essentially the same for different possible choices of \(S_0\)}.
The results for the different $R$ shown in Fig.~\ref{fig:extrapolation} confirm that our model describes the data also for values of $M$ much smaller than the region used in fixing $S_0,a$.

\begin{figure}[]
\includegraphics[width=1\columnwidth]{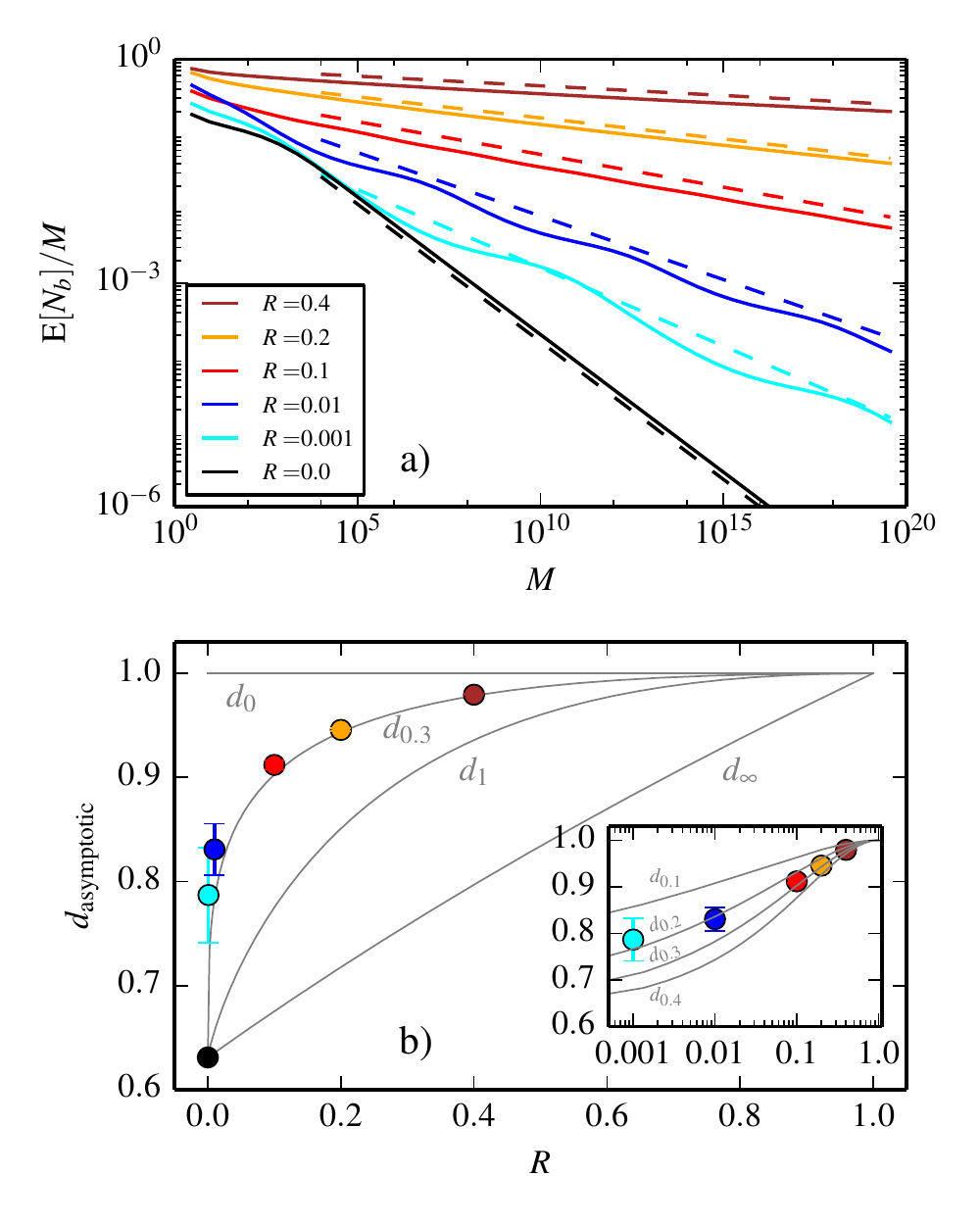}
\caption
   {
      (Color online)
      Asymptotic scaling of the classical prediction.
      a) Extrapolation of $\mathrm{E}\left[N_b\right]$ of the curves in Fig.~\ref{fig:extrapolation}a without the shift (i.e. ${a=1}$).
      $R$ increases from the black (bottom) to the brown (top) curve. 
      The dashed lines follow the asymptotic scaling $M^{d_\mathrm{asymptotic}-1}$ with \(d_\mathrm{asymptotic}=\langle\frac{\ln N(M)-\ln N(M')}{\ln M-\ln{M'}}\rangle\), where the average is computed over all possible combinations \(\left\{(M,M'): M'<M\land (M,M')\in\{3^{10},3^{11},\dots,3^{40}\}^2\right\}\).
      In b) we show these dimensions together with their standard deviations.
      In grey are analytically obtained $d_q$ curves, see Eq.~\eqref{eq:dq_analyt}.
      Note, that for fixed $R$ and $q\in\mathbb{R}^+$, $d_0\geq d_q\geq d_\infty$ \cite{Ott:Book2002}.
   }
\label{fig:asymptotics}
\end{figure}

The quantitative success of our sampling model motivates us to consider its behavior in the semiclassical limit~$M\rightarrow\infty$. 
In Fig.~\ref{fig:asymptotics}a we plot Eq.~(\ref{eq:expectation}) in an increased range of $M=1/\varepsilon$ values using for the different $R$ curves the value of $S_0$ used in Fig.~\ref{fig:extrapolation}b (estimated from the quantum data). 
We again see the presence of log-periodic oscillations (see Ref.~\cite{Sornette:PR1998} for a review on log-periodicities in fractals).
While these oscillations play a major role in $d_{\mathrm{loc}}$ and for the range of $M$ accessible in the quantum computations, they do not hinder the estimation of an asymptotic dimension $d_{\mathrm{asymptotic}}$ as the general trend of the curves.
For \(R=0\) the results confirm that for $d_{\mathrm{asymptotic}}=d_0^{(R=0)}$ and that our sampling correctly accounts for the usual fractal Weyl law. 
More interestingly, $d_{\mathrm{asymptotic}}<1$ for all $R$, strongly suggesting that the non-trivial scaling $d_{\mathrm{loc}}$ observed in the quantum data are {\it not} finite~$M$ effects but persist in the asymptotic limit.
This implies that the fractal Weyl law of partially open systems does not follow the classical prediction $d_0=1$. 
Comparing the $R$ dependence of $d_{\mathrm{asymptotic}}$ with the analytical calculated $d_q$ indicates that roughly $q\approx 0.3$ can be used to estimate this scaling. 

\section{Discussions}

In summary, we have investigated the statistical properties of long-lived eigenstates of chaotic systems with partial absorption.
Contrary to the naive prediction obtained combining the classical fractal dimension~$d_0$ and the fractal Weyl law, we observe numerically that for a broad range of parameters the number of long-lived states grows sub-linearly with the dimensionality~$M$ of the Hilbert space.
We explain these observations by considering a model which is consistent with the quantum to classical correspondence in the semiclassical limit $M\rightarrow\infty$ but that accounts for an undersampling of the classical measure at finite $M$.
This undersampling combined with the strong spatial fluctuations of the multifractal classical measure is responsible for the effective reduction of the available phase-space volume for long-lived eigenstates.

Our most surprising finding is that even the asymptotic ($M\rightarrow\infty$) scaling of the number of long-lived states differ from the $d_0=1$ prediction.
This result is compatible with our numerical simulations at least for a broad range of cutoffs in $|\nu|$.
In any practical situation the limit $M\rightarrow \infty$ is not achievable and the relevance of our model is that it allows to understand the effective dimension observed quantum mechanically.
We have also considered alternative models in which we assume that the eigenstates concentrate exclusively on the phase-space regions with highest measure $\mu$, or on boxes with a \(\mu\)-dependent probability that we compute without including a sampling process.
None of these models is able to reproduce the quantum observations, see Appendix~\ref{sec:app:alternatives}.

Our results are not limited to area-preserving maps or to systems with localized absorbing regions.
Instead, they should be visible in systems of any dimension for which a multifractal spectrum of a classical measure $\mu$ appears.
In $n$-dimensional systems the long-lived resonances are expected to concentrate in $n$-cuboids  of volume \(h\) and the sampling has to be carried out on a hyperplane intersecting these volumes.
In general partially-absorbing systems there is no analogous fully open system ($R\rightarrow0$ for the cases above).
In these cases, the scaling of Weyl's law  is expected to follow the dimension of the regions of highest probability\footnote{The $d_0$ of the peaks is different from $d_q$ for $q\rightarrow \infty$.} for small $M$ (analogous to $d_0$ of the $R=0$ case in the examples above) and show a transition towards the dimension $d_\mathrm{asymptotic}$ of the undersampled $\mu$ (for the large $M$ case). 
The multi-fractality in the localization of wave functions can appear also without a clear classical counterpart, e.g. in the case of phase transitions in disordered systems ~\cite{Evers:RMP2008,Martin:PRE2010}.  It would be interesting to investigate whether the number of such states  show similar dependencies on system size as the ones we observed.

\begin{acknowledgments}
   We thank  A. B\"acker,  R. Ketzmerick, M. K\"orber, and T. T\'el for insightful discussions and suggestions.
\end{acknowledgments}

\FloatBarrier

\appendix

\section{Analytical $D_q$ in the baker map}\label{ap:baker}
Assume a ternary baker map as defined in Fig.\ref{fig:sketches} with reflectivities $R_\mathrm{L}$ in the left, $R_\mathrm{M}$ in the middle and \(R_\mathrm{R}\) in the right vertical strip.
In this system it is possible to calculate the number of necessary iterations to get an exact result up to some minimum $\varepsilon_\mathrm{min}$.
The size of features (stripes) in the ternary baker map after \(n\) iterations is at least $1/3^n$.
If $\varepsilon_\mathrm{min}$ is smaller than that value then the box-counting will pick up the features as areas and $D_0$ will approach 2.

Let \(R_\mathrm{L,M,R}\in\left[0,1\right]\).
After the first iteration the intensity in the top horizontal third is $R_\mathrm{L}$, the central strip has an intensity of $R_\mathrm{M}$, and the bottom third carries $R_\mathrm{R}$.
The total intensity is
\begin{equation}
   I_\mathrm{tot,1}=R_\mathrm{L}+R_\mathrm{M}+R_\mathrm{R}.
\end{equation}
The next iteration gives the following total intensity:
\begin{eqnarray}
   I_\mathrm{tot,2}=
      &R_\mathrm{L}^2+R_\mathrm{L}R_\mathrm{M}+R_\mathrm{L}R_\mathrm{R}&+\nonumber\\
      &R_\mathrm{M}R_\mathrm{L}+R_\mathrm{M}^2+R_\mathrm{M}R_\mathrm{R}&+\\
      &R_\mathrm{R}R_\mathrm{L}+R_\mathrm{R}R_\mathrm{M}+R_\mathrm{R}^2&\nonumber
\end{eqnarray}
and so forth.
At the $n$-th step we obtain the total intensity
\begin{eqnarray}
   I_{\mathrm{tot},n}&=&\sum_{k=0}^n\sum_{l=0}^{n-k}\binom{n}{k,l}R_\mathrm{L}^kR_\mathrm{M}^lR_\mathrm{R}^{n-k-l}\\
   &=&\left(R_\mathrm{L} +R_\mathrm{M} +R_\mathrm{R} \right)^n\nonumber
   \label{eq:def:total_intensity_after_n}
\end{eqnarray}
Here we used the fact that the number of strips with a certain combination of reflectivities \(R_\mathrm{L}^kR_\mathrm{M}^lR_\mathrm{R}^{n-k-l}\) is given by the multinomial coefficient \(\binom{n}{k,l}:=\frac{n!}{k!l!(n-k-l)!}\).

Now we focus on the multifractal dimension spectrum \eqref{eq:def:renyi_dimensions} of the distribution generated by this map.
First, since the measure is constant in the horizontal direction, we can restrict ourselves to a vertical line and calculate the partial dimensions \(d_q\) along this line.
Assume this line is covered by disjoint segments of length \(\varepsilon=3^{-n}\).
The dimensions of the full distributions are then simply obtained from
\begin{equation}
   D_q=1+d_q
\end{equation}
with
\begin{equation}\label{eq:def:dq}
   d_q=\frac{1}{1-q}\lim_{n\rightarrow \infty}\frac{\ln\sum_{i=1}^{3^n}\mu_i^q}{n\ln 3}.
\end{equation}
Here $\mu_i$ is the intensity on the \(i\)-th segment normalised by $I_{\mathrm{tot},n}$, so that 
\begin{equation}
   \sum_{i=1}^{3^n}\mu_i=1
\end{equation}
Or, again using multinomial factors:
\begin{equation}
   \sum_{k=0}^n\sum_{l=0}^{n-k}\binom{n}{k,l}\mu_{n;k,l}=1
\end{equation}
with 
\begin{equation}\label{eq:def:mu_nkl}
   \mu_{n;k,l}=\frac{R_\mathrm{L}^kR_\mathrm{M}^lR_\mathrm{R}^{n-k-l}}{I_{\mathrm{tot},n}}
\end{equation}
With that we can write the sum in \eqref{eq:def:dq} as
\begin{eqnarray}
   \ln\sum_{i=1}^{3^n}\mu^q_i=\\
   \ln\sum_{k=0}^n\sum_{l=0}^{n-k}\frac{\binom{n}{k,l}\left(R_\mathrm{L}^kR_\mathrm{M}^lR_\mathrm{R}^{n-k-l}\right)^q}{\left(R_\mathrm{L} +R_\mathrm{M} +R_\mathrm{R}\right)^{nq}}=\\
   \ln\frac{\left(R_\mathrm{L}^q+R_\mathrm{M}^q+R_\mathrm{R}^q\right)^n}{\left(R_\mathrm{L} +R_\mathrm{M} +R_\mathrm{R}\right)^{nq}}=\\
   n\ln\frac{\left(R_\mathrm{L}^q+R_\mathrm{M}^q+R_\mathrm{R}^q\right)}{\left(R_\mathrm{L} +R_\mathrm{M} +R_\mathrm{R}\right)^{q}}
\end{eqnarray}
Thus we obtain for $d_q$
\begin{equation}\label{eq:dq_analyt}
   d_q=\frac{1}{1-q}\lim_{n\rightarrow \infty}\frac{n\ln\frac{\left(R_\mathrm{L}^q+R_\mathrm{M}^q+R_\mathrm{R}^q\right)}{\left(R_\mathrm{L} +R_\mathrm{M} +R_\mathrm{R}\right)^{q}}}{n\ln 3}.
\end{equation}
This is independent of $n$ so we can drop the limit and get
\begin{equation}
   d_q=\frac{\ln\frac{\left(R_\mathrm{L}^q+R_\mathrm{M}^q+R_\mathrm{R}^q\right)}{\left(R_\mathrm{L} +R_\mathrm{M} +R_\mathrm{R}\right)^{q}}}{(1-q)\ln 3}.
\end{equation}
The limit $q\rightarrow1$ can be obtained using L'H\^opital's rule. 
\begin{align}
   d_1:=& \lim_{q\rightarrow 1} \frac{1}{1-q}\frac{\ln \sum_{i=1}^{3^n}\mu^q_i}{n\ln 3}\nonumber\\
   \Rightarrow d_1=&\frac{\ln\left(R_\mathrm{L}+R_\mathrm{M}+R_\mathrm{R}\right)}{\ln{3}}\\
   &-\frac{R_\mathrm{L}\ln R_\mathrm{L}+R_\mathrm{M}\ln R_\mathrm{M}+R_\mathrm{R}\ln R_\mathrm{R}}{\left(R_\mathrm{L}+R_\mathrm{M}+R_\mathrm{R}\right)\ln(3)}.\nonumber
\end{align}
Asymptotically $d_q$ goes to
\begin{equation}
   \lim_{q\rightarrow \infty} d_q=\frac{\ln\left(R_\mathrm{L} +R_\mathrm{M} +R_\mathrm{R}\right)-\ln\left(R_\mathrm{max}\right)}{\ln 3},
\end{equation}
where \(R_\mathrm{max}:=\max\left\{R_\mathrm{L},R_\mathrm{M},R_\mathrm{R}\right\}\).

\section{Quantised maps}\label{sec:app:quantum_maps}

The matrix representation of the propagator of the quantised version of the baker map defined in Fig.~\ref{fig:sketches} is
\begin{equation}
   \mathcal{M}_\mathrm{baker}=
   \mathcal{U}_\mathrm{baker}
   \cdot
   \begin{bmatrix}
      \mathbb{1}_{M/3}&0&0\\
      0&\sqrt{R}\mathbb{1}_{M/3}&0\\
      0&0&\mathbb{1}_{M/3}\\
   \end{bmatrix}
\end{equation}
with
\begin{equation}\label{eq:def:quantum_baker}
   \mathcal{U}_\mathrm{baker}=
   \mathbf{F}^{-1}_M\cdot
   \begin{bmatrix}
      \mathbf{F}_{M/3}&0&0\\
      0&\mathbf{F}_{M/3}&0\\
      0&0&\mathbf{F}_{M/3}\\
   \end{bmatrix}
\end{equation}
where \(\mathbf{F}_{M;m,n}=\frac{1}{\sqrt{M}}\exp\left[-\frac{2\pi\mathrm{i}}{M}(n+1/2)(m+1/2)\right]\) are Fourier transformations from position to momentum representation and \(\mathbf{F}^{-1}_M\) denotes the inverse transformation.
For an explanation of a quantisation scheme see e.g.~\cite{Balazs:AnnPhys9891}.

Analogously, the quantised cat map has the following form:
\begin{equation}\label{eq:def:quantum_cat}
   \mathcal{M}_\mathrm{cat}=
   \left[\alpha_{M;m,n}\right]_M\cdot
   \begin{bmatrix}
      \mathbb{1}_{M/6}&0&0\\
      0&\sqrt{R}\mathbb{1}_{M/2}&0\\
      0&0&\mathbb{1}_{M/3}\\
   \end{bmatrix}
\end{equation}
where \(M\) is even and the entries are~\cite{Kuznetsov:PD2000}
\begin{equation}
   \alpha_{M;m,n}=\frac{1}{\sqrt{M}}\exp\left[ \frac{2\pi \mathrm{i}}{M}\left( m^2 -mn + n^2\right)\right].
\end{equation}

\section{Derivation of \(\mathrm{E}[N_b]\)}\label{sec:app:undersampling}

Starting from the general equation \eqref{eq:expected_value_general} for the expected number of accepted boxes here we show how to arrive at the formula for the baker map \eqref{eq:expectation}.

The \(i\)-th box is found with a probability \(p_i=1-(1-\mu_i)^S\).
Analogous to \eqref{eq:def:mu_nkl} we can restrict ourselves to \emph{different} boxes and define \(p_{n:k,l}:=1-(1-\mu_{n;k,l})^S\).
The expected value of the number of found boxes is
\begin{equation}
   \mathrm{E}\left[N_b\right]=\sum_{i=1}^{3^n}p_i=\sum_{k=0}^n\sum_{l=0}^{n-k}\binom{n}{k,l}p_{n;k,l}.
\end{equation}
Plugging in the expression for \(\mu_{n;k,l}\) we get
\begin{align}
   \label{eq:expectationvalue}
   &\mathrm{E}\left[N_b\right]=\\&\sum_{k=0}^n\sum_{l=0}^{n-k}\binom{n}{k,l}\left(1-\left(1-\frac{R_\mathrm{L}^kR_\mathrm{M}^lR_\mathrm{R}^{n-k-l}}{\left(R_\mathrm{L} +R_\mathrm{M} +R_\mathrm{R}\right)^{n}}\right)^S\right).\nonumber
\end{align}
Equation~\eqref{eq:expectation} follows from setting \(R_\mathrm{L,R}=1\), \(R_M=R\), carrying out the inner sum, and renaming the indices.

\section{Discussion of alternatives to sampling}\label{sec:app:alternatives}

Here we want to show that taking into account only the regions of phase-space with the highest measures by a cutoff in the measure can not reproduce the scaling observed in the quantum maps.

First, let us consider a fixed cutoff.
In this case it is easy to see that for any value of \(\mu_\mathrm{cutoff}=const.\) there exists a \(n^*\) with \[\mu_{n;k,l}<\mu_\mathrm{cutoff},\,\forall n>n^*.\]
Therefore \(d_0\xrightarrow{n\rightarrow\infty}0\).

The second case is an \(n\)-dependent cutoff value \(\mu_{\mathrm{cutoff},n}=\frac{I_\mathrm{cutoff}}{I_{\mathrm{tot},n}}\) with constant \(I_\mathrm{cutoff}\).
In this case if all \(R_\mathrm{L,M,R}<1\) there also exists a \(n^*\) with \(\mu_{n;k,l}<\mu_{\mathrm{cutoff},n},\,\forall n>n^*.\)
In other words: \(d_0\xrightarrow{n\rightarrow\infty}0\).
An exception here arises if \(R_\mathrm{L,R}=1\) and \(R_\mathrm{M}\in\left[0,1\right)\).
Then, for all \(0<I_\mathrm{cutoff}<1\) after a number of iterations \(n^*\) all middle thirds fall below the cutoff \(d_0\xrightarrow{n\rightarrow\infty}\ln 2/\ln 3\) from above.

So, by associating the number of long-lived eigenstates of boxes above a given threshold we obtain an effective dimension contradicting our numerical observations.
This indicates that the (large) regions with low measure $\mu$ still contain substantial portions of the eigenstates.

The third alternative we want to discuss is the case where each box contributes to the long-living phase-space volume proportionally to its intensity
\begin{equation}
   p_{n;k,l}\propto I_{n;k,l}=\mu_{n;k,l}I_{\mathrm{tot},n}. 
\end{equation}
As before we identify \(M=3^n\).
In this context we need to limit the contribution of each box \(p_{n;k,l}=\min\{1,\alpha I_{n;k,l}\}\) and compute \(\mathrm{E}\left[N_b\right]=\sum_{k=0}^n\sum_{l=0}^{n-k}\binom{n}{k,l}p_{n:k,l}\).
The factor \(\alpha\) plays a similar role as the cutoff in the sense that for \(\alpha\rightarrow\infty\) (\(\alpha\rightarrow 0\)) we count all (zero) boxes.
This method reproduces the correct scalings in the cases \(R_\mathrm{L,M,R}=1\), and \(R_\mathrm{L,R}=1\),\(R_\mathrm{M}=0\).
However, it fails to reproduce the increase of the effective dimension with $R$ and $M$ and the oscillating behaviour of \(d_\mathrm{loc}\) with \(|\nu|_\mathrm{cutoff}\). 

\bibliography{bibl}

\begin{thebibliography}{26}
\bibitem[{foo()}]{foot1}
\bibinfo{note}{The choice of $M$ as a multiple of $3$ respects the partition of
  the baker map in three regions. In the cat map, the quantisation scheme
  requires $M$ to be even and therefore we use $M_\mathrm{cat}=3^n+1$ with
  $k\in\mathbb{N}$.}
\end{thebibliography}

\end{document}